\documentclass[%
 reprint,
 amsmath,amssymb,
aps,
pra,
]{revtex4-2}

\usepackage{graphicx}
\usepackage{dcolumn}
\usepackage{amsmath,amssymb,amsfonts}%
\usepackage{bm}

\begin{document}

\preprint{APS/123-QED}

\title{Strain engineering of magnetic anisotropy in the kagome magnet Fe$_3$Sn$_2$}%

\author{Deli Kong}
 \altaffiliation[Also at ]{Department of Materials Science and Engineering, Southern University of Science and Technology,  Shenzhen, 518055, China}
\author{Andr\'as Kov\'acs}%
\email{a.kovacs@fz-juelich.de}
 \altaffiliation[Also at ]{Institute of Technical Physics and Materials Science, HUN-REN Centre for Energy Research, Budapest, 1121, Hungary} \email{a.kovacs@fz-juelich.de}
 \author{Rafal E. Dunin-Borkowski}
\affiliation{Ernst Ruska-Centre for Microscopy and Spectroscopy with Electrons, Forschungszentrum J\"ulich, J\"ulich, 52425 Germany
}%

\author{Michalis Charilaou}
\affiliation{ Department of Physics, University of Louisiana at Lafayette, Lafayette, 70504 USA} \email{michalis.charilaou@louisiana.edu}%

\author{Markus Altthaler}
\author{Lilian Prodan}
\author{Vladimir Tsurkan}
\altaffiliation[Also at ]{Institute of Applied Physics, Moldova State University, Chisinau, MD 2028 Moldova}
\author{Istv\'an K\'ezsm\'arki}
\affiliation{Experimental Physics V, University of Augsburg, Augsburg, 86135 Germany}

\author{Xiadong Han}
\email{hanxd@sustech.edu.cn}
\affiliation{Department of Materials Science and Engineering, Southern University of Science and Technology,  Shenzhen, 518055, China}

\author{Dennis Meier}
\affiliation{Department of Materials Science and Engineering, and Center for Quantum Spintronics, Department of Physics, NTNU Norwegian University of Science and Technology, Trondheim, 7491, Norway}

\date{\today}%

\begin{abstract}
The ability to control magnetism with strain offers innovative pathways for the modulation of magnetic domain configurations and for the manipulation of  magnetic states in materials on the nanoscale. Although the effect of strain on magnetic domains has been recognized since the early work of C. Kittel, detailed local observations have been elusive. Here, we use mechanical strain to achieve reversible control of magnetic textures in a kagome-type Fe$_3$Sn$_2$ ferromagnet without the use of an external electric current or magnetic field \emph{in~situ} in a transmission electron microscope at room temperature. We use Fresnel defocus imaging, off-axis electron holography and  micromagnetic simulations to show that tensile strain modifies the structures of dipolar skyrmions and switches their magnetization between out-of-plane and in-plane configurations. We also present quantitative measurements of magnetic domain wall structures and their transformations as a function of strain. Our results demonstrate the fundamental importance of anisotropy effects and their interplay with magnetoelastic and magnetocrystalline energies, providing new opportunities for the development of strain-controlled devices for spintronic applications.
\end{abstract}

                    
\maketitle


\section{Introduction}
Increasing interest in quantum materials is based in part on the synergy between strongly correlated electron states, topology and magnetism, which can lead to unconventional physical  properties. Such systems include superconductors, topological semiconductors, Weyl semi-metals, quantum spin liquids and two-dimensional (2D) materials, in which quantum effects are manifested over a wide range of energy and length scales \cite{Keimer2017}. Magnetism on a kagome lattice, a 2D network of corner-sharing triangles, provides a versatile platform for investigating the interplay between magnetic phenomena, topology and electronic correlations \cite{Kiyohara2016,Liu2019,Ortiz2020, Yin2022}. Kagome layers of 3d transition metals, stacked in perpendicular directions, offer an interesting realization of this concept. For example, kagome-type Fe$_3$Sn$_2$ exhibits a wide range of intriguing properties, such as flat bands near the Fermi energy \cite{Lin2018}, massive Dirac fermions \cite{Ye, Kang2020}, and a large anomalous Hall effect \cite{Kida_2011, Ye, Zhang2021, Schilbert2022}, combined with a high Curie temperature (670~K) \cite{Fenner2009}. The material has a centrosymmetric rhombohedral structure of Fe-Sn bilayers, which alternate with Sn layers along the crystallographic $c$-axis (Fig.~\ref{fig1}a, and  Supplemental Material Fig. S1), resulting in competing uniaxial ($K_u$) and shape anisotropies with a quality factor of $<$1 at room temperature. In thin films of Fe$_3$Sn$_2$, a competition between perpendicular magnetic anisotropy and shape anisotropy can lead to both type-I and type-II magnetic bubble formation \cite{Kezsmarki2021}, with diverse helicities \cite{Du2024}. 
In such structures, the spin texture of the bubbles is non-uniform, introducing additional surface spin twists and internal degrees of freedom. Similar spin structures have also been observed in the 2D van-der-Waals-type ferromagnet Fe$_5$GeTe$_2$ \cite{Parkin2024}, and are typically termed unconventional bubbles \cite{Parkin2024} or dipolar skyrmions \cite{Du2024}. We use the latter name in the present work. The possibility of driving dipolar skyrmions with an electric current \cite{Du2021} and of detecting them using anisotropic magnetoresistance \cite{Tang2023} makes Fe$_3$Sn$_2$ of particular interest for spintronic applications.

In addition to applied electric currents and external magnetic fields, there is great potential in introducing and controlling strain in a material system, which adds another degree of freedom when engineering device architectures with desired properties.
In spintronics, strain engineering is used to introduce magneto-elastic coupling in magnetostrictive ferromagnets \cite{Wang}.  Strain has also been applied to skyrmion-hosting materials \cite{Shibata,Hu,Hatton_2022,Du_2023}. However, few experimental studies have been carried out to investigate local changes in strain-induced magnetic states on the nanoscale. Recently, Liu et al. \cite{Liu2024} reported strain-induced reversible motion of skyrmions using inhomogenous uniaxial compressive strain. In our previous work, we presented strain-induced-hardening in ferromagnetic Ni thin films \cite{Kong2023} in the presence of uniform tensile strain. 
Recent advances in development of straining devices \cite{Chengpeng,Dongwei} now allow \emph{in situ} tensile straining experiments with unprecedented resolution and precision. 

Here, we show how magnetic texture in Fe$_3$Sn$_2$, which comprises dipolar skyrmions, can be controlled by means of mechanical strain at room temperature, without the need for an electric current or external magnetic field. We select this kagome compound because it is expected to display a large strain effect as it undergoes magnetic reorientation below room temperature, driven by a change in magnetocrystalline anisotropy from easy-axis-type to easy-plane-type \cite{Malaman1978, Heritage2020, He2022}. We use Lorentz transmission electron microscopy (TEM) and off-axis electron holography to directly observe and quantify strain-induced magnetization rotation in single crystalline Fe$_3$Sn$_2$. We show that dipolar skyrmions  merge into a periodic stripe domain structure, before the out-of-plane magnetic field gradually rotates to an in-plane direction parallel to the applied strain direction. In this state, large magnetic domains are separated by bow-tie domain walls. On removing the strain, the magnetization returns to its initial out-of-plane arrangement. The observed magnetization rotation is thought to result from the original magnetic easy-axis anisotropy being over-ridden by strain-induced anisotropy, which favors in-plane magnetic spin alignment. Such precise and reversible control of a magnetic state using mechanical strain opens new possibilities for the design of advanced spintronic devices that exploit the intricate interplay between different anisotropies.

\section{Results and Discussion}

\begin{figure*}%
\centering
\includegraphics[width=0.9\textwidth]{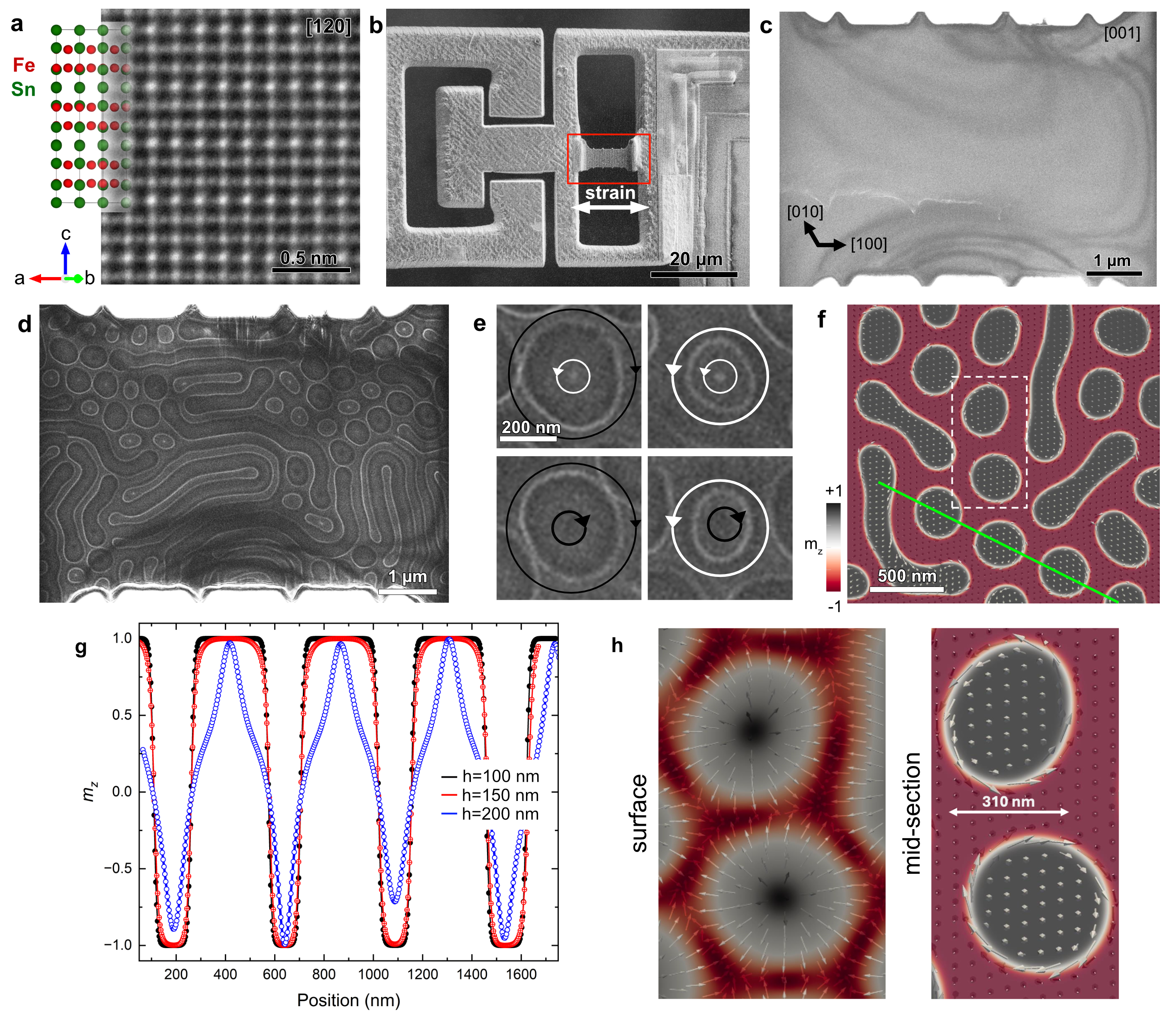}
\caption{Tensile straining of Fe$_3$Sn$_2$.
(a)~Unit cell and an atomic resolution HAADF STEM image of Fe$_3$Sn$_2$ viewed along the [120] crystallographic direction.
(b)~Secondary electron SEM image of an Fe$_3$Sn$_2$ TEM lamella mounted on the mechanical tensile device.
(c)~Bright-field TEM image of the single crystal Fe$_3$Sn$_2$ lamella. 
(d)~Fresnel defocus image of Fe$_3$Sn$_2$, showing stripe domains and dipolar skyrmions of type-I. The defocus value is 0.8 mm. 
(e)~Selected images showing the four basic magnetic field rotation diversities of dipolar skyrmions. White and black circles indicate counterclockwise and clockwise in-plane directions, respectively. 
(f)~Micromagnetic simulation displayed in the form of the mid-section at $z=h/2$.
(g)~Line profile along the green line in (f) showing the magnetization ($m_z$) at the surface and in a bulk section of the dipolar skyrmions. 
\textbf{h} Section of micromagnetic simulation of two dipolar skyrmions marked in (f), showing N\'eel and Bloch spin textures at the surface and bulk. Note the opposite rotation of the magnetization in the mid-section.
}
\label{fig1}
\end{figure*}

Figure~\ref{fig1}b shows an electron-transparent Fe$_3$Sn$_2$ specimen mounted in a miniature straining device, which acts in the horizontal direction marked by a white arrow. The Fe$_3$Sn$_2$ specimen is single crystalline, with no visible structural defects (Fig.~\ref{fig1}c). The viewing direction is parallel to [001] and the horizontal [100] axis is parallel to the strain direction.
The energetically-favorable domain state in a thin Fe$_3$Sn$_2$ film comprises stripe-like domains, which can be controlled by a magnetic field applied in the kagome plane. (See Fig.~S2 in the Supplemental Material). In order to form dipolar skyrmions, a magnetic field of 0.53~T is applied parallel to the magnetic easy axis, \emph{i.e.}, perpendicular to the kagome plane, using the conventional objective lens of the microscope. The applied field is then removed to image the magnetic state at remanence, as shown in Fig.~\ref{fig1}d for a Fresnel defocus image of dipolar skyrmions coexisting with stripe domains. Analysis of the image contrast shows that the dipolar skyrmions have diverse helicities \cite{Kong2023}. Magnified images are shown in Fig.~\ref{fig1}e. The magnetic field arrangement in each dipolar skyrmion has a two-fold rotation sense, \emph{i.e.}, clockwise or counterclockwise in the inner and outer regions. (See Fig.~S3 in the Supplemental Material). A balance between ferromagnetic exchange and dipole-dipole interactions results in a hybrid Bloch-N\'eel spin twist, which varies through the thickness of the specimen in a three-dimensional manner. The black and white contrast in the Fresnel defocus images (Fig.~\ref{fig1}d,~e) can be used to identify the dipolar skyrmions and their direction of rotation. In analogy to conventional magnetic bubbles,  dipolar skyrmions in Fe$_3$Sn$_2$ exhibit type-I and type-II configurations. The latter configuration contains two domain walls in the peripheral region. The schematic representation of the different magnetic structures and their Fresnel defocus images can be found in Fig. S4 in the Supplemental Material.
Figure~\ref{fig1}f shows the result of a micromagnetic simulation of Fe$_3$Sn$_2$, which reproduces the observed magnetic texture.  (See the Methods section for details). Circular dipolar skyrmions with clockwise and counterclockwise rotations and randomly-oriented stripe domains are visible. The simulation also reproduces the Bloch-type to N\'eel-type magnetization rotation of the dipolar skyrmions, which can be followed in the variation of the $m_z$ magnetization component with depth, as shown in Fig.~\ref{fig1}g. Figure~\ref{fig1}h illustrates the N\'eel-type arrangement on the surfaces and the Bloch-type arrangement in the mid-sections \cite{Kezsmarki2021} of two dipolar skyrmions.

\subsection{\emph{In~situ} tensile straining }\label{straining}

\begin{figure*}[ht]%
\centering
\includegraphics[width=0.7\textwidth]{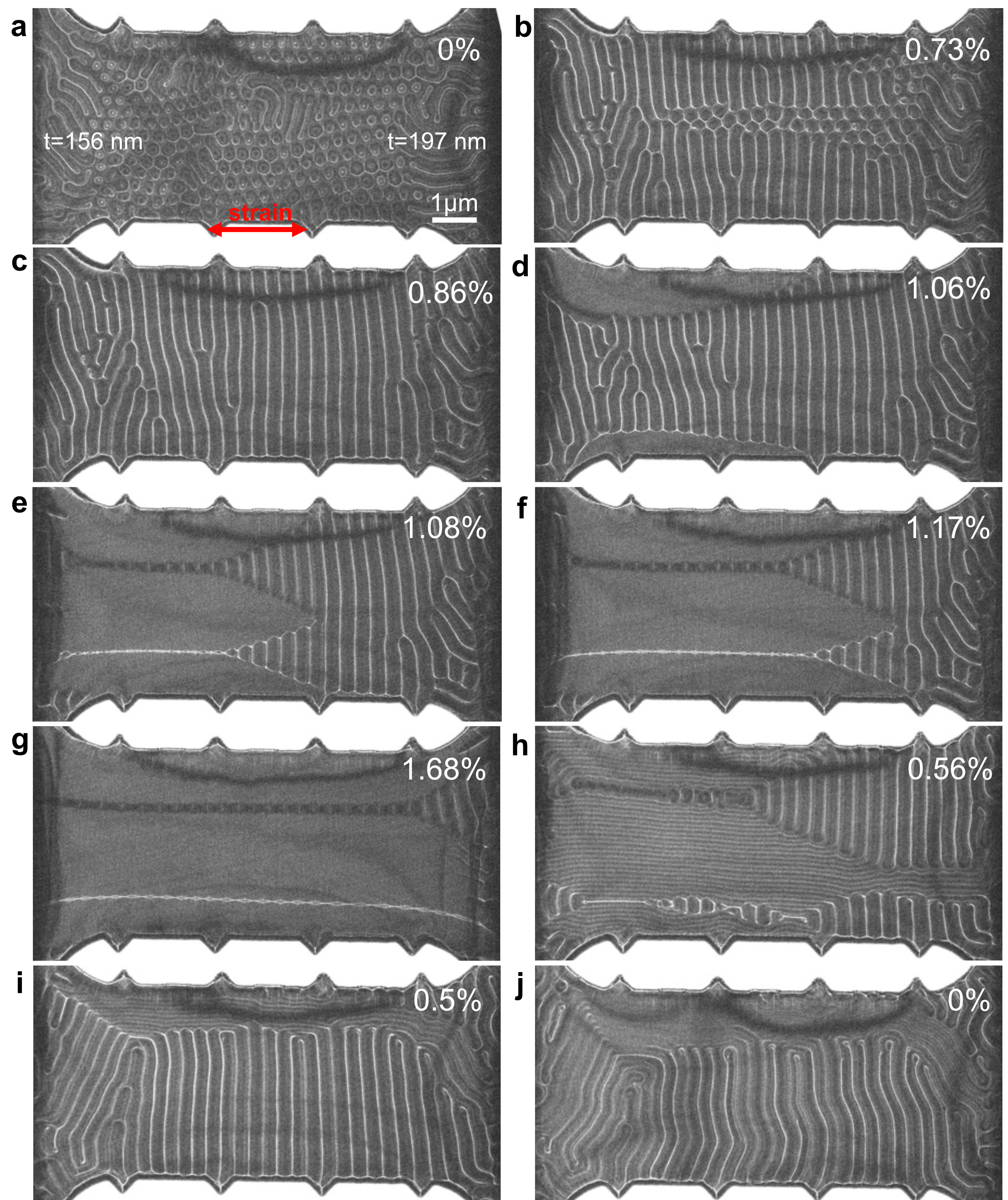}
\caption{Fresnel defocus images of  \emph{in~situ} (a-f) tensile straining and (g-j) releasing of Fe$_3$Sn$_2$ recorded at magnetic remanence. The strain is applied along the horizontal direction, marked using an arrow in (a). The defocus value is 0.8 mm.}
\label{fig2}
\end{figure*}

Figure~\ref{fig2} shows Fresnel defocus images of the effect of tensile strain on the magnetic state of Fe$_3$Sn$_2$ under magnetic field free conditions. The lateral dimensions of the specimen are approximately 10~$\mu$m~$\times~$4~$\mu$m (width $\times$ height), while its thickness varies from 156 to 197~nm within the field of view in Fig.~\ref{fig2}a. A total of 229 dipolar skyrmions was counted. In the presence of a strain of approximately 0.8~\%, the dipolar skyrmions were replaced by stripe domains oriented perpendicular to the strain direction (Fig.~\ref{fig2}b,~c). Above a strain of 1~\%, uniform domains formed near the specimen edge (Fig.~\ref{fig2}d) and grew to form linear domain walls parallel to the strain direction (Fig.~\ref{fig2}e). Closure of the vertical domain walls then resulted in the formation of horizontal domain walls (Fig.~\ref{fig2}e-g). The horizontal domain walls, which were formed by vertical narrowing of the original stripe pattern, preserved the periodic contrast variation of the stripes, with a period of 190$\pm$18~nm.
At a maximum strain of 1.68~\% (Fig.~\ref{fig2}g), the specimen contained large domains separated by straight domain walls. 
When the strain was released (Fig.~\ref{fig2}h-j), vertical stripe domains formed again, enlarging as the strain was decreased to zero. The transition from a vertical domain structure to large domains and back in Figs~\ref{fig2}e-h resembles a zipper mechanism. (See Movie~S1 in the Supplemental Material).
Dipolar skyrmions were not observed in the fully relaxed state in this specimen. However, their re-formation was observed in samples that were highly strained and had passed through the fracture point (see Fig.~S5 in the Supplemental Material), suggesting that a small amount of residual strain may remain once the externally applied stress is released. The fine balance between different anisotropies in Fe$_3$Sn$_2$ results in a strong thickness dependence of the magnetic structure. For thicknesses below 100~nm, magnetocrystalline anisotropy is not strong enough to support dipolar skyrmions and stripe domains with out-of-plane magnetization. (See Fig.~S5 in the Supplemental Material). Furthermore, position control of dipolar skyrmions can be realized by using mechanical strain in the presence of a magnetic field (see Movie~S2 in the Supplemental Material), with the positions of dipolar skyrmions changing directionally in the presence of strain and returning to their original positions when strain is released.

The \emph{in~situ} experiment described in Fig. \ref{fig2} reveals three primary strain-induced processes in the Fe$_3$Sn$_2$ thin film: (i)~the formation of a regular vertical stripe pattern by the gradual elimination and merging of dipolar skyrmions up to a strain of 1~\%; (ii)~the formation of large domains with in-plane magnetization separated by straight domain walls (1-1.7~\%); (iii)~recovery of the vertical stripe pattern upon strain release.

\begin{figure*}%
\centering
\includegraphics[width=0.8\textwidth]{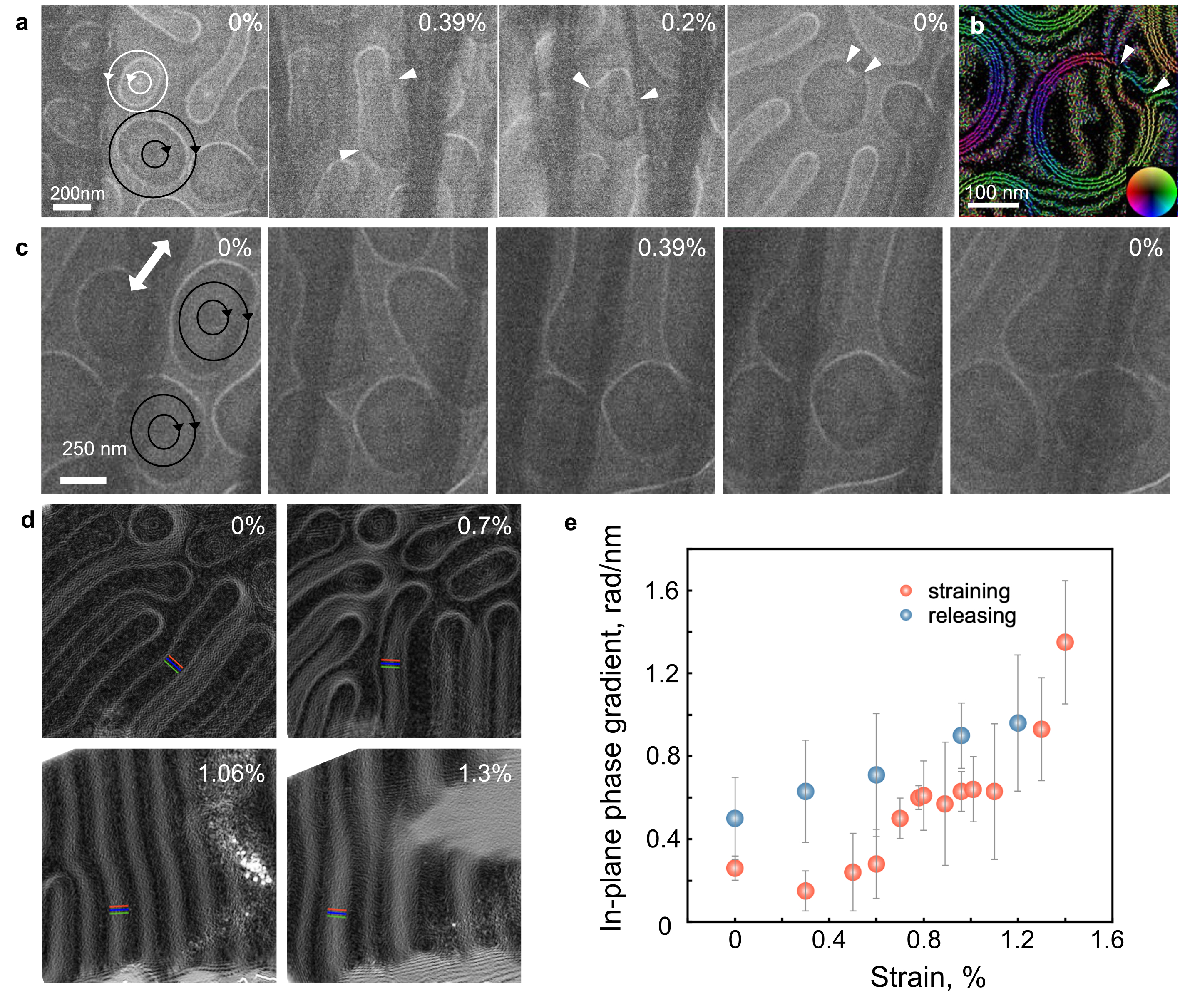}
\caption{Lorentz TEM and off-axis electron holography of strain induced effects in Fe$_3$Sn$_2$.
(a)~Merging of dipolar skyrmions with opposite helicities at moderate strain. Triangular arrowheads mark domain walls. The defocus value is 0.6 mm.
(b)~In-plane magnetic induction map of the resulting type-II dipolar skyrmion, measured using off-axis electron holography. The direction of the projected in-plane magnetic induction is visualized according to the color wheel. (c)~The anti-parallel field alignment at the peripheries of two dipolar skyrmions is robust against skyrmion merging at high strain.  The defocus value is 0.6 mm.
(d)~Formation of stripe domains perpendicular to the strain direction. The marked region was used to estimate the in-plane phase gradient values plotted in (e), which reveal changes in projected in-plane magnetic induction as the strain increased (red circles) and then released (blue circles). The error bars correspond to the standard deviations of the phase gradient. }
\label{fig3}
\end{figure*}

At moderate strains (up to 0.7~\%), the dipolar skyrmions merge \emph{via} two characteristic mechanisms, which are presented in Fig.~\ref{fig3}a-c. In the first case, clockwise and counterclockwise rotation of the magnetization rotation of adjacent dipolar skyrmions is associated with parallel field alignment at their peripheries. In the presence of strain, the dipolar skyrmions merge to form an elliptical shape, whose opposite ends maintain the original magnetization rotation. The outer edge of the newly-formed dipolar skyrmion contains two domain walls, which form a type-II structure and are marked by arrowheads in Fig.~\ref{fig3}a. No core structure is observed in the merged particle, suggesting the presence of a hybrid texture \cite{Du2024}, in which the upper and lower surfaces of the skyrmion have opposite helicities. As the strain is released, the two domain walls move closer together, until a small segment of about 50~nm remains between them. Figure~\ref{fig3}b shows a magnetic induction map of such a type-II structure measured using off-axis electron holography. At the two domain walls, the in-plane field directions are head-to-head and tail-to-tail. A similar merging process of skyrmions with opposite helicities to form a particle with a complex spin structure is presented in Fig.~S6 in the Supplemental Material. In the second case (Fig. \ref{fig3}c), clockwise rotation of the magnetization of adjacent dipolar skyrmions is associated with anti-parallel field alignment at their peripheries. This configuration remains robust against merging up to higher strain levels.  

We analyzed the internal magnetic field alignment in the vertical stripe domains. Figure~\ref{fig3}d shows the gradient of the electron optical phase measured using of-axis electron holography at different strain levels.  As the electron optical phase is sensitive to in-plane field components projected in the incident electron beam direction, field rotation inside the Fe$_3$Sn$_2$ specimen can be inferred from the phase gradient measured at the same location as a function of strain. The in-plane phase gradient in Fig.~\ref{fig3}e increases significantly as the strain is increased to 1.30~\%, which is consistent with the magnetization in the sample rotating from out-of-plane to in-plane. 

The most striking and unexpected observations of the strain-release cycle in the Fe$_3$Sn$_2$ thin film presented in Fig.~\ref{fig2} are that the configuration changes to domain walls that are perpendicular to the strain axis before aligning with the strain axis, and that the initial out-of-plane magnetization direction rotates away from the original magnetocrystalline anisotropy axis and aligns in-plane in the presence of strain before returning reversibly to the out-of-plane direction when the strain is released.  

\subsection{Tuning of magnetocrystalline anisotropy energy \emph{via} the magnetoelastic effect}

\begin{figure*}%
\centering
\includegraphics[width=0.8\textwidth]{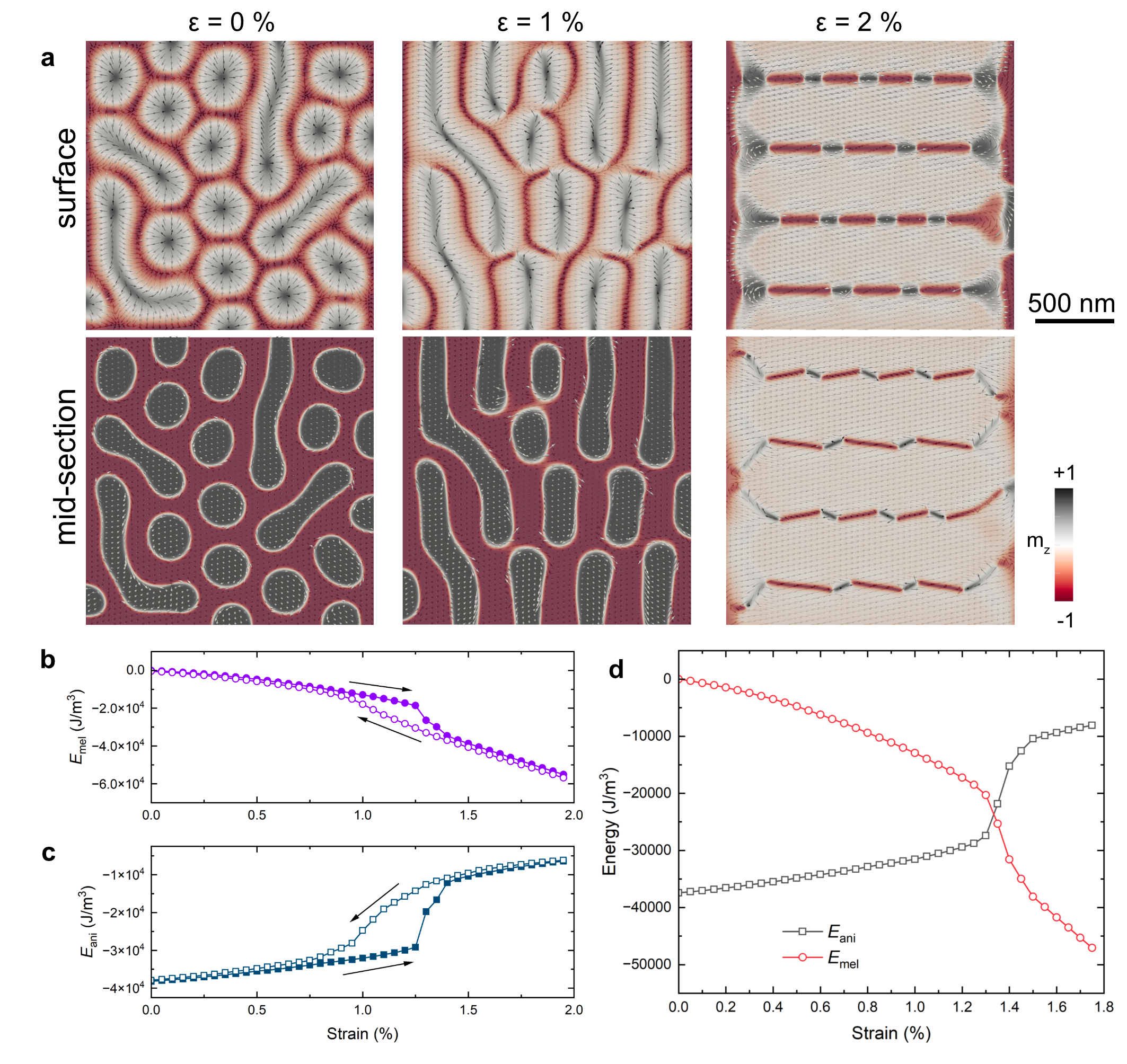}
\caption{Micromagnetic simulation results of Fe$_3$Sn$_2$.
(a)~Upper and middle sections, showing N\'eel and Bloch arrangements as a function of strain (0, 1 and 2\%). 
(b,c)~Calculated anisotropy energy ($E_{ani}$) and magnetoelastic energy ($E_{mel}$), revealing hysteretic behavior during strain loading and unloading.
(d)~The onset of in-plane magnetization occurs when the magnetoelastic energy density becomes greater than the magnetocrystalline anisotropy energy.
}
\label{fig4}
\end{figure*}

Figure~\ref{fig4} shows micromagnetic simulations of the magnetic state of the Fe$_3$Sn$_2$ lamella at different strains. The first row from the top shows the state of the surface of the lamella, while the second row shows the state at half of the lamella thickness ($z~=~h/2$).

In these simulations, the first-order magnetoelastic constant $B_1$ was used as a fitting parameter to achieve a match with the experimental results. The value of $B_1$ determines the strain at which the magnetoelastic energy $E_\mathrm{mel}$ becomes larger than the original magnetocrystalline uniaxial anisotropy energy $E_\mathrm{ani}$. By comparing the simulations with the experimental results, in which in-plane rotation of the magnetization was observed at $\epsilon\approx 1.3\%$, a value for $B_1$ of -3100$\pm$100 J/m$^3$ was obtained.
Significantly, we find that the simulations reproduce all of the key experimental features: (i)~the initial state that contains stripe domains and dipolar skyrmions with both helicities; (ii)~merging and rotation of the stripe domains and dipolar skyrmions perpendicular to the strain axis with increasing tensile strain along the $x$ axis; (iii)~abrupt rotation of the magnetization to the in-plane direction at a critical strain of $\epsilon\approx 1.3\%$; (iv)~return of the system to a state comparable to the original state upon decreasing the strain, while exhibiting hysteretic behavior.

Based on the simulations and on the analysis of the domain wall structures, we propose that the hysteretic behavior results from configurational anisotropy \cite{Cowburn1998}, \emph{i.e.}, it is associated with the topology of the magnetization textures and the processes by which the local magnetization twists to eliminate stripe domains.

The formation of stripe domains is unexpected. Conventionally, the magnetization would be expected to gradually turn towards the strain direction and not to form domains perpendicular to it. Our analysis here reveals that the formation of stripe domains perpendicular to the strain axis satisfies exchange energy, magnetoelastic energy and magnetostatic energy, in contrast to the penalty in magnetocrystalline anisotropy energy. At a critical strain of 1.3\%, there is a drastic change in the energy balance, with $E_\mathrm{ani}$ increasing by 50\%, while at the same time $E_\mathrm{mel}$ decreases by 50\% (see Fig.~4d).

\subsection{Domain wall width analysis}

\begin{figure*}[ht]%
\centering
\includegraphics[width=1.0\textwidth]{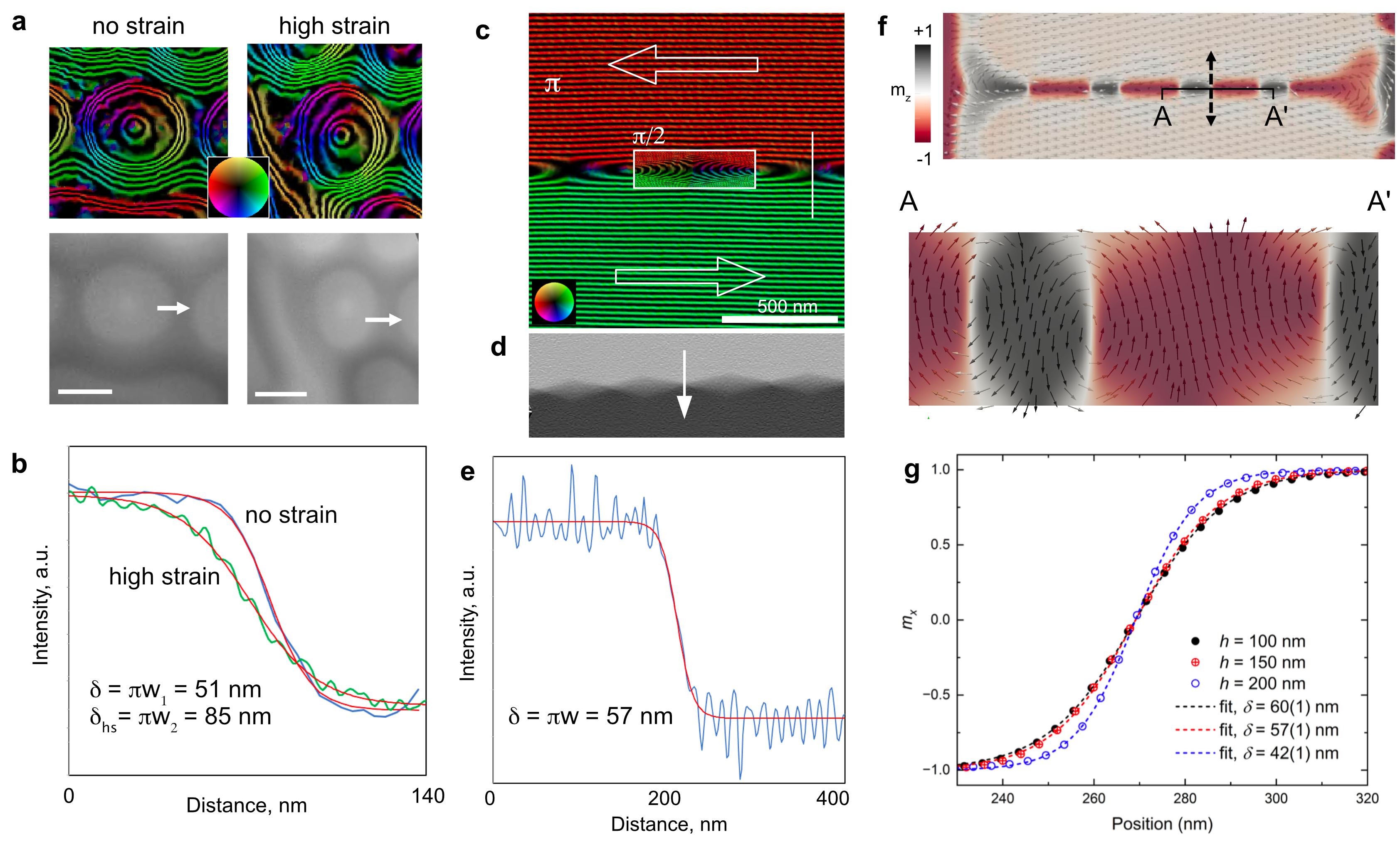}
\caption{Domain wall width analyses of Fe$_3$Sn$_2$. 
(a)~Magnetic induction maps and phase shift images of dipolar skyrmions imaged in relaxed (no strain) and high strain conditions. The phase contour spacing is 2$\pi$/6 radians. The scale bar is 200~nm.
(b)~Line profiles extracted from the phase shift images in (a) across domain walls at the positions of the arrows. The line profiles are fitted using \textit{tanh} functions. The measured domain wall width increases from 51$\pm$9 to 85$\pm$9~nm with strain for dipolar skyrmions.
(c)~Magnetic induction map of in-plane magnetic domains imaged at high strain. Arrows and colors highlight the 180$^{\circ}$ alignment of the in-plane magnetic field directions. The contour spacing in the rectangular box is increased to $\pi$/2 radians to visualize the induction lines of the bow-tie domain wall.
(d,e)~Differential of the phase shift and line profile used to determine a a value for the domain wall width of 57$\pm$9~nm. 
(f)~Micromagnetic simulation of in-plane magnetic domains with a bow-tie wall showing the surface (top). The cross section marked (A-A') reveals the depth variation of the field direction in the bow-tie domain wall. The double-head arrow indicates the location of the linescan of $m_z$ shown in (g). The fitting reveals the wall width variation.
}
\label{fig5}
\end{figure*}

The off-axis electron holography results and micromagnetic simulations allow the domain wall structure in thin films of Fe$_3$Sn$_2$ to be analyzed, in particular for the bow-tie domain wall that forms at high strain levels. Figure~\ref{fig5} shows experimental and theoretical analyses of domain walls in dipolar skyrmions and bow-tie domain walls. Domain wall width measurements were carried out for dipolar skyrmions in the initial and highly strained conditions (Fig.~\ref{fig5}a). Figure~\ref{fig5}b shows phase profiles extracted across the edges of dipolar skyrmions marked by arrows in Fig.~\ref{fig5}a. Each profile was fitted using a hyperbolic tangent function, according to the expression $y$~=~$y_0$~+~$a$~tanh(($x$~-~$x_0$)/$w$), where $y_0$, $a$, $x_0$ and $w$ are constants obtained from the fit. The width of the domain wall can be defined as $\delta$~=~$\pi w$. In this way, the domain wall width for Fe$_3$Sn$_2$ dipolar skyrmions was determined to be 51$\pm$9~nm for the initial condition and 85$\pm$9~nm for the strained condition. Figure~\ref{fig5}c shows a magnetic induction map of a bow-tie domain wall measured using off-axis electron holography. The parallel contour lines and colors confirm the 180$^{\circ}$ alignment of the projected in-plane field. The inset to Fig.~\ref{fig5}c shows part of the magnetic induction map with a phase contour spacing of $\pi$/2 radians, which helps to visualize the field lines at the domain wall. The decrease in contour spacing in the middle section suggests out-of-plane magnetic field alignment. The term ``bow-tie domain wall'' is inspired by the well-known cross-tie domain wall structure \cite{Hubert}(page 225), which comprises 90$^{\circ}$ segments of circular and cross Bloch lines. In the case of a cross-tie domain wall, the in-plane magnetic induction is uniform and the density of the induction lines is constant. Experimental measurements of a conventional cross-tie wall in a soft ferromagnetic alloy using off-axis electron holography were presented in  Ref.\cite{Kovacs2018}(Fig.5, p.74). In contrast, the bow-tie wall presented here contains segments where the magnetic field is oriented in the out-of-plane direction (See Fig. S4 in the Supplemental Material). By using the phase gradient (Fig.~\ref{fig5}d), the domain wall width was measured to be 57$\pm$9~nm (Fig.~\ref{fig5}e), which is similar to the value measured for unstrained dipolar skyrmions. 
Figure \ref{fig5}f shows a top view and a cross-section of a bow-tie domain wall extracted from the micromagnetic simulation along the line A-A'. The arrows show an alternation of out-of-plane orientation and rotation within the segments.

In our micromagnetic simulations of dipolar skyrmions, we used the expression presented above to fit profiles of the $z$ component of the magnetization at different $z$ positions, since the domain wall is not uniform through the thickness of the sample. In the center of the lamella ($z$~=~$h$/2), the width is 43$\pm$2~nm, whereas between the center and the surface ($z$~=~3$h$/4) it is 55$\pm$2~nm, yielding an average domain wall width of 49$\pm$2~nm, in close agreement with the experimental result presented in Fig.~\ref{fig5}b. In contrast, the domain wall thickness in the strained state (Fig.~\ref{fig5}f) is larger in the center of the lamella (60$\pm$2~nm) and becomes smaller towards the surfaces (42$\pm$2 nm), yielding an average domain wall width of 51$\pm$2~nm.

\section{Conclusion}

Our study of mechanical-strain-induced effects on magnetic states in thin films of kagome-type Fe$_3$Sn$_2$ using \emph{in~situ} magnetic imaging in the transmission electron microscope and a MEMS-based straining nanodevice reveals the elimination and merging of dipolar skyrmions, the formation of regular stripe domains and rotation of the preferred out-of-plane magnetization direction to in-plane as the strain is increased, while the system stays in the elastic regime. We identify two dipolar skyrmion merging mechanisms under low strain conditions and the formation of a bow-tie domain wall, which comprises both in-plane and out-of-plane field components, in a highly strained condition.
 Micromagnetic simulations reveal a hysteretic strain effect on the anisotropy and magnetoelastic energies and show that the formation of stripe domains perpendicular to the strain axis satisfies exchange, magnetoelastic and magnetostatic energies. Our observations highlight the potential for strain-controlled magnetism in nanomagnetic devices, and underscore the need for local studies of the interplay between anisotropic effects, magnetoelastic energies and magnetocrystalline properties for advancing practical applications in magnetic technology.

\section{Methods}

\subsection{Specimen preparation and straining}.
\emph{In~situ} deformation in the TEM was performed using a Bestron (Beijing) Science and Technology Co., LTD \emph{in~situ} double tilt TEM specimen straining holder (INSTEMS-MET)\cite{Jianfei}. Focused Ga ion beam sputtering in a dual beam scanning electron microscope (ThermoFisher Helios NanoLab 460F1) was used to position and thin an Fe$_3$Sn$_2$ single crystal that had been prepared using a chemical transport method \cite{Kezsmarki2021} in the rectangular frame of a micro-electromechanical systems (MEMS) chip. Markers were fabricated at the edge of the lamella for strain measurement, as shown in Fig.~1b. Strain was applied to the specimen in the Bestron holder using the MEMS chip, which contains a lead zirconate titanate (PZT) actuator for displacement control. In this setup, hooks catch a T-head to approximately realize a uniaxial tensile test of the lamella. 

\subsection{Magnetic imaging}
Images of magnetic domain walls and projected in-plane magnetic induction were recorded and quantified using Fresnel defocus imaging and off-axis electron holography. Off-axis electron holograms were recorded using a spherical aberration corrected TEM (ThermoFisher Titan 60-300) operated at 300~kV. Magnetic-field-free conditions were realized by switching off the conventional microscope objective lens and using the transfer lens of the aberration corrector for imaging. Fresnel defocus images and off-axis electron holograms were recorded on a 4k $\times$ 4k pixel direct electron counting detector (Gatan K2 IS). The typical biprism voltage was 90~V, which corresponded to a holographic interference fringe spacing of 3.02 nm and holographic interference fringe contrast measured in vacuum of 60\%. Image analysis was performed using Gatan Microscopy Suite and HoloWorks software. Total phase shift information was extracted using a standard Fourier transform method. The total phase shift provides information about local variations in both electrostatic and magnetic potential. Since the lamella is a single crystal with negligible thickness variations, it was assumed that the electrostatic contribution to the signal is constant and that any variations in phase away from the sample edge are magnetic in origin. Magnetic induction maps were generated by in the form of contours obtained from the recorded phase images and colors determined from their gradients. 

\subsection{Micromagnetic simulation}
The Fe$_3$Sn$_2$ sample was modeled as a ferromagnet with exchange stiffness $A$, magnetoelastic coupling $B_1$, saturation magnetization $M_\mathrm{s}$ and first-order and second-order perpendicular magnetocrystalline anisotropies $K_1$ and $K_2$, respectively.

The magnetoelastic energy density of a hexagonal lattice is \cite{Koch2004}
\begin{align}
E_\mathrm{mel}=&B_1\left(m_x^2\epsilon_{xx}+m_y^2\epsilon_{yy}+m_xm_y\epsilon_{xy}\right)-B_2m_z^2\epsilon_{zz}\notag \\
&-B_3m_z^2\left(\epsilon_{xx}+\epsilon_{yy}\right)+B_4\left(m_ym_z\epsilon_{yz}+m_xm_z\epsilon_{xz}\right)
\end{align}
To the best of our knowledge, these constants are not known for Fe$_3$Sn$_2$. However, in our experiments tensile strain was only applied along the $x$ axis. Therefore, the only non-vanishing terms in the above equation are the $B_1m_x^2\epsilon_{xx}$ and $-B_3m_z^2\epsilon_{xx}$ terms. Given that the latter term has the same $m_z$ dependence as the first-order uniaxial anisotropy, we set $B_3~=~0$ and considered only the first order magnetoelastic constant $B_1$ in our approximation. The energy density of our model then takes the form
\begin{align}
E=& \sum_{i=x,y,z}A\left(\nabla m_i\right)^2 -K_1\left(m_z\right)^2 -K_2\left(m_z\right)^4\notag \\
&+B_1 \left(m_x\right)^2\epsilon_{xx} -\mu_0M_\mathrm{s}\left(\mathbf{H}+\frac{1}{2}\mathbf{H}_\mathrm{dip}\right)\cdot \mathbf{m}
\end{align}
where $m_{i=x,y,z}$ are components of the magnetization unit vector $\mathbf{m}=\mathbf{M}$/$M_\mathrm{s}$, $\mathbf{H}$ is the external field vector. $\mathbf{H}_\mathrm{dip}$ is the magnetic field vector due to long-range dipole-dipole interactions and $\epsilon$ is the mechanical strain.

The values of the saturation magnetization and the two uniaxial anisotropy constants were measured experimentally and found to be $M_\mathrm{s}~=~5.66\times10^5$~A/m, $K_1~=~5.2\times10^4$~J/m$^3$ and $K_2~=~6.5\times10^3$~J/m$^3$, respectively. Our value for $M_\mathrm{s}$ is closer to that reported in Ref.~\cite{Hou2017} compared to that reported in Refs~\cite{Du2023,Du2024}. Our value of the first-order uniaxial anisotropy constant is in good agreement with Refs~\cite{Du2020,Du2023,Du2024}. For the exchange stiffness, we estimated a value by comparing the Curie temperature of Fe$_3$Sn$_2$ with that of pure Fe, and obtained a value of $A~=~12\times10^{-12}$~J/m, which is higher than the value reported in Refs~\cite{Du2020,Du2023,Du2024} and lower than the value estimated in Ref.~\cite{Hou2017} from the experimentally-determined domain wall width. Our approach for estimating $A$ provides a more accurate estimate because the Curie temperature can be measured more accurately than the domain wall width, which depends on the imaging resolution and sample thickness (cf. \cite{Kong2023}).

In our sample, the exchange length is $\delta~=~\sqrt{2A/\mu_0 M_\mathrm{s}^2}~=~7.7$~nm. In the simulation, the sample had dimensions of 2000~nm $\times$ 2000~nm $\times$ 200~nm, which is comparable to the dimensions of the sample used in our experiments. The cell size was 4~nm $\times$ 4~nm $\times$ 4~nm, and the micromagnetic simulations were performed using Mumax3 \cite{Mumax3}. Tests using other cell sizes were performed to confirm numerical stability. The simulation protocol was to start the system with an initial state of $\mathbf{m}~=~\left(0,0,1\right)$ and numerically integrate the Landau-Lifschitz-Gilbert (LLG) equation $\partial_t\mathbf{m}~=~-\gamma\left(\mathbf{m}\times\mathbf{H}_\mathrm{eff}\right)+\alpha\left(\mathbf{m}\times\partial_t\mathbf{m}\right)$, where $\mathbf{H}_\mathrm{eff}~=~-\partial E/\partial \mathbf{m} \mu_0M_\mathrm{s}$ is the effective field. The LLG equation was integrated for 20~ns until the system reached equilibrium in zero field and for zero strain. Then, the tensile strain $\epsilon_{xx}$ was increased gradually from 0 to 2\% in steps of 0.05\%. At each step, the LLG equation was integrated numerically for 2~ns to allow the system to reach a new equilibrium.

\begin{acknowledgments}
The authors are grateful for funding from the European Research Council under the European Union's Horizon 2020 Research and Innovation Programme (Grant No.~856538, project ``3D~MAGiC''), to the Deutsche Forschungsgemeinschaft (DFG, German Research Foundation) \emph{via} projects No.~405553726 (TRR270), No.~492547816 (TRR260) and No.~49254781 (TRR 360), to the Research Council of Norway (No.~263228) and Centres of Excellence funding scheme  (No.~262633, ``QuSpin''), to the ``111'' project (DB18015), to the National Key R\&D Program of China (2021YFA1200201) and to Project No.~ANCD20.80009.5007.19 (Moldova). N.S.~Kiselev is acknowledged for stimulating discussions.
\end{acknowledgments}

%

\end{document}